# Confinement controlled bend instability of three-dimensional active fluids

Pooja Chandrakar[1,2], Minu Varghese[1], S.Ali Aghvami[1], Aparna Baskaran[1], Zvonimir Dogic[2,1] and Guillaume Duclos[1]

[1] Department of Physics, Brandeis University, Waltham, MA 02474
[2] Department of Physics, University of California at Santa Barbara, Santa Barbara, CA 93106

Spontaneous growth of long-wavelength deformations is a defining feature of active fluids with orientational order. We investigate the effect of biaxial rectangular confinement on the instability of initially shear-aligned 3D isotropic active fluids composed of extensile microtubule bundles and kinesin molecular motors. Under confinement, such fluids exhibit finite-wavelength self-amplifying bend deformations which grow in the plane orthogonal to the direction of the strongest confinement. Both the instability wavelength and the growth rate increase with weakening confinement. These findings are consistent with a minimal hydrodynamic model, which predicts that the fastest growing deformation is set by a balance of active driving and elastic relaxation. Experiments in the highly confined regime confirm that the instability wavelength is set by the balance of active and elastic stresses, which are independently controlled by the concentration of motors and non-motile crosslinkers.

Active fluids are suspensions of motile constituents [1,2]. For anisotropic active fluids, energy-consuming rod-like units generate active stresses that are proportional to the local nematic order parameter. A consequence of such stresses is the intrinsic bend instability of a uniformly aligned suspension [3-7]. Orientational fluctuations generate self-amplifying deformations, which destroy long-range order, leading to a steady-state characterized by autonomous chaotic flows [8-13]. Boundaries and geometrical confinement can transform such inherently chaotic dynamics [14-23]. For example, three-dimensional annular channels convert chaotic flows into long-ranged coherent flows capable of transporting materials on meter scales [23]. However, the exact mechanisms by which boundaries transform locally generated active stresses to potentially produce macroscale flows remain under investigation. Two parallel planar walls is a simple geometry to investigate boundary effects. Previous measurements of the hydrodynamic length scale in chaotic 3D active fluids – including dense bacterial suspensions [8,24], living liquid crystals [10], and reconstituted cytoskeletal networks [9] - were performed in such geometries. To understand how boundaries affect the structure of an active fluid, it is essential to systematically increase the smallest confinement dimension [25,26].

In this letter, we demonstrate that aligned 3D extensile active fluids are intrinsically unstable, and we study how boundaries determine the emergent length scale of the instability. Minimal active hydrodynamic theories predict that unconfined bulk fluids are always unstable at non zero activity, and that the fastest-growing deformation has an infinite wavelength [3,7]. Active hydrodynamic flows equally amplify orientational fluctuations at all length scales. However, the higher elastic costs associated with short-wavelength modes imply that long-wavelength deformations are more unstable. Both boundary and substrate friction limit the range of unstable modes [27-31], the latter case being experimentally studied in 2D active nematics coupled to a viscous oil layer [32,33]. Except in some peculiar geometries [34,35], confined active nematics are unstable only when the active stress is large enough to overcome elastic and viscous damping [4,6,19], and the fastest-growing mode has a finite wavelength [4]. We assemble 3D isotropic active fluids composed of biopolymers and molecular motors and study how biaxial channels with a rectangular cross-section determine the nature of their orientational instability. Experiments demonstrate that i) size of the channel determines the length scale of the growing deformations; ii) bulk dynamics cannot be reached as the active fluid structure is determined by the boundaries for all experimentally realizable geometries; iii) boundaries set the dominant instability plane; iv) weakening confinement leads to larger wavelengths and faster dynamics, and (v) the instability switches from an initially exponential to linear growth regime. The first four experimental findings are consistent with the linear stability analysis of a minimal hydrodynamic model.

We study an isotropic active fluid composed of microtubules (MT) bundled by a non-specific depletant [9,36]. The suspension is driven away from equilibrium by clusters of kinesin motors, which hydrolyze Adenosine 5′-triphosphate (ATP) to step-along and drive apart adjacent antiparallel microtubules. Such extensile units push fluid along both directions of their long axis while pulling fluid along their short axis. Hence, they exert dipolar extensile stresses and belong to the symmetry class of pusher-type swimmers [1]. The active fluid composed of dilute MT-bundles (~0.1% volume fraction) forms a continuously reconfiguring network that powers autonomous flows. Importantly, MT-based active fluids can be assembled on milliliter scales with dynamics that persist for hours, allowing us to study the influence of boundaries on macroscopic scales [37].

We confined MT-based active isotropic fluids within microfabricated channels, whose heights ($H$, z-axis) were either smaller or equal to their widths ($W$, y-axis) (**Fig. 1E**). Similar to passive rod-like colloids [38], flowing MT aligned them along the channel's length (x-axis). After the external flow ceased, we observed the growth of periodic in-plane bend deformations with spatially uniform wavelength (x-y plane, **Fig. 1A, 1B, SI materials and methods, Movies S1, S2**).
Since the MT-bundles are extensile, fluctuations about the aligned state generate unbalanced internal forces that act on the surrounding fluid. Further, when these fluctuations consist of bend deformations, the flows generated by extensile activity amplify the initial deformations, giving rise to an instability that is well known as the generic bend instability [4]. Instability required both ATP and a critical concentration of molecular motors. In the highly confined regime ($W$= 2.5 mm, $H$= 100 μm), the instability was entirely suppressed for motor cluster concentrations less than 5 nM (**Fig. S1, Movie S3**). Similar in-plane instabilities have been observed in kinesin-free suspension of microtubules aligned by a magnetic field, where the extensile stress was generated by filament growth [39]. Intriguingly, different out-of-plane motor-driven buckling has also been reported for suspensions of longer microtubules and lower motor concentrations [40,41]. Out-of-plane buckling suggests a more solid-like character of the MT network, a hypothesis that was independently tested by rheological measurements [42].

To quantify the instability, we measured both the temporal evolution of the in-plane velocity v and the nematic director **n**[cos(θ), sin(θ)], where θ is the angle between the local bundle alignment and the channel's flow direction (x-axis). Both the velocity and nematic director field exhibited periodic patterns from which we extracted the wavelength and the amplitude of the fastest growing mode (**Fig. S2, SI**). The amplitude first grew exponentially as expected from the linear stability analysis. Intriguingly, after a well-defined time, it transitioned to a linear growth, which persisted until saturation (**Fig. 1C**). The transition from exponential to linear growth was observed over a wide range of motor concentrations (**Fig. S3**). Since the MT suspension is dilute, we did not observe the creation of disclination loops, in contrast to 3D active nematics [43,44]. Instead, the instability melts the initial nematic order into an isotropic active fluid [9].

The temporal evolution of the MT alignment was quantified by the alignment parameter $q=<\cos(2\theta)>$ (**Fig. 1D**). During the exponential growth, MT-bundles were well aligned ($q \sim 1$) along the channel's long axis. As the nematic director buckled, the bundles increasingly pointed along the y-axis, which led to a decreasing $q$. Eventually, the alignment parameter q became negative, reaching a minimum $q \sim -0.5$. At that point, growth slowed and the bundles were on average aligned along the y-axis. The resultant state, although not as ordered as the initial state, was itself unstable, leading to the growth of deformations along the x-direction. Similar to 2D nematics [32], a cascade of instabilities orthogonal to each other ensued (**Fig. S4**). The wavelength of the successive instabilities increased over time until the fluid reached the previously described isotropic structure and lost all memory of its initial nematic order. The wavelength increase could be due to an increase of the elastic constant from the coarsening of the MT-bundles, and/or a decrease of the degree of ordering of the successive aligned states prior to each instability cycle (**Fig. S4E**).

The instability can be described by the linearized equations of 3D active nemato-hydrodynamics for an extensile fluid (**SI**). These equations recover features of the previously studied generic hydrodynamic instability [3]. In the absence of boundaries, the fastest growing mode at any nonzero activity corresponds to pure-bend deformations and the instability wavelength is infinite. Introducing no-slip boundaries changes several features of the instability. First, the active stress must be larger than a critical value for deformations to spontaneously grow (**Fig. S1**), as previously reported for confined 2D active nematics [4,6,19,45]. Second, above that threshold, there is a finite range of unstable wavelengths. Third, in confined 3D active nematics both twist-bend and splay-bend modes are unstable, with the former dominating in initially aligned suspension due to their faster growth (**SI**). To further investigate the last prediction, we used confocal microscopy, which confirmed that the 3D instability was not pure-bend, as the in-plane director was not uniform at different z-planes (**Fig. S5**). Moreover, in-plane bend deformations of the MT-bundles grew faster than splay deformations (**Fig. S6**).

The instability is triggered by self-amplifying orientational fluctuations that destabilize the initially shear-aligned nematic order. In bulk active fluids, orientation fluctuations are isotropic. When confined, the linear stability analysis revealed that the unstable twist-bend modes are given by $\boldsymbol{x} \cdot (\boldsymbol{k} \times \boldsymbol{\delta n}_\perp) = \frac{\delta n_z}{W} - \frac{\delta n_y}{H}$, where $\boldsymbol{k}$ is the wavevector of the unstable mode, and $\boldsymbol{\delta n}_\perp = (\delta n_y, \delta n_z)$ is a sinusoidal perturbation perpendicular to x, the direction of the initial director field (**SI**). If $H \ll W$, the unstable mode consists predominantly of $\delta n_y$. This is consistent with experiments: spontaneous growth of deformations was most conspicuous in the image (x-y) plane, which is orthogonal to the strongest confinement direction (z-axis) (**Fig. 1A**).

So far, we described the instability of initially shear-aligned active fluids in the highly confined regime (*H*=100 μm, *W*=3 mm). Next, we systematically changed the channel's width and height to determine how confinement size controls the wavelength selection mechanism. Confocal images of labeled microtubules confirmed that the bundles were initially aligned with the channel direction for all geometries (**SI Movie 5**). We started by increasing the channel height (*H*, z-direction) from 100 μm to 3 mm while keeping its width (*W*, y-direction) fixed at 3 mm (**Movie S6**). The instability wavelength and the associated growth rate strongly depended on the channel height, increasing rapidly for small channels (*H* < 1 mm) before plateauing for the larger channels (**Fig. 2A, Fig. S7A**). Repeating these experiments for smaller widths showed that a combination of both the width and height sets the fastest growing mode (*W*=1 mm, **Fig. 2A**).

Linear stability analysis of a minimal hydrodynamic model reveals that the orientationally ordered state is unstable to deformations when the confinement length scale $L = \left(\frac{1}{W^2} + \frac{1}{H^2}\right)^{-1/2}$ is larger than an active lengthscale $\ell_\alpha = 2\pi \sqrt{\frac{2\eta D_R \kappa}{\alpha(S_0 + \xi)}}$ (**SI**). Here, $\alpha$ is the magnitude of the extensile force dipole generated by the MT-bundles, $S_0$ is the magnitude of initial order, $\xi$ is the flow alignment parameter, $\eta$ is the viscosity, $D_R$ is the rotational diffusion constant, and $\kappa$ is the nematic elasticity. The wavelength of the fastest growing mode, $\lambda$, depends on the channel dimensions as:

$\lambda = 2 \left[ \frac{1}{\pi l_\alpha L} - \frac{1}{L^2} \right]^{-1/2}$ (**SI**). As the channel goes from a rectangular to a square cross-section, the theoretical scaling of the instability wavelength with the channel size agrees qualitatively with the experimental observations (**Fig. 2B**). Furthermore, both experiments and theory demonstrate that the instability wavelength is mainly determined by the smaller channel dimensions and does not significantly change with increasing the larger channel dimension. For example, if the channel height is fixed at 50 µm, increasing the channel width from 125 µm to 500 µm does not influence the instability wavelength, nor the growth rate (**Fig. 2C, Fig. S7B, Movie S7**). Finally, for channels where $H,W \gg \ell_\alpha$, the linear stability analysis predicts that $\lambda \propto \sqrt{\frac{h}{\sqrt{1+h^2}}} \cdot \sqrt{W}$, where $h = H/W$ is the aspect ratio of the channel's cross-section.

While both theory and experiments exhibit similar qualitative trends, there are quantitative inconsistencies. For instance, fitting the hydrodynamic theory to the experimental data predicts that $\ell_\alpha \sim 400$ µm, suggesting that the instability should be suppressed for channel sizes that are significantly larger than what is observed experimentally. This discrepancy might be because the MT suspension we study is in isotropic phase. Alternativity, the hydrodynamic model might require additional terms to describe the instability in the strong confinement regime ($H,W<400$ µm).

Many existing experimental realizations of active matter – from shaken granular rods to multicellular tissues – are two dimensional. Motility in 2D often requires contact with a fluid or a solid substrate. Frictional interactions with a substrate screen any long-range hydrodynamic flows. For example, the instability wavelength of 2D MT-based active nematics is finite due to the viscous damping with the adjacent oil layer [32]. In comparison, the finite wavelength observed here in 3D active fluids is induced by confinement. In bulk, the most unstable mode is pure-bend with an infinite wavelength. Confinement in the direction of initial order restricts the maximum allowed wavelength so that the most unstable mode is a pure-bend mode with a wavelength equal to the system size. For confinements perpendicular to the direction of initial order, the presence of no-slip boundaries constraints the velocity and director fields such that deformations can no longer be pure-bend. Consequently, splay and twist deformations are unavoidable in the channel geometry. Long-wavelength bend deformations are suppressed because the active driving cannot overcome the viscous and elastic costs that come with the mandatory splay and twist deformations. Short-wavelength modes have a larger fraction of bend compared to the splay and twist deformations; hence activity destabilizes these modes more than the longer-wavelength modes. Shortest wavelength modes are again stabilized by elasticity and viscous dissipation. Consequently, this competition between active and dissipative processes implies that the most unstable mode has a finite wavelength.

So far, we demonstrated that the channel size controls both the growth rate and the wavelength of the instability. Next, we control the instability wavelength by tuning the intrinsic material properties of an active fluid, such as the activity or the network elasticity. Linear stability analysis predicts that the fluid becomes unstable at length scales larger than $l_\alpha \propto \sqrt{\frac{\kappa}{\alpha}}$, where $\kappa$ is the sample elasticity, and $\alpha$ is the activity coefficient (SI) [12,46]. For strong confinements (W=3 mm, H=100 µm), the wavelength slightly decreases with increasing ATP concentration (Fig. 3A). It is often assumed that ATP controls the magnitude of the active stress. However, this dependency does not necessarily account for the biochemistry behind kinesin stepping: processive motors take multiple steps on a microtubule before unbinding. After most steps, the motor waits for the ATP molecule to bind before taking the next step [47]. Clusters of processive motors can bind to multiple microtubules and therefore have a dual role, acting both as passive cross-linkers and generators of active stresses [48]. With decreasing ATP, kinesin dwell time increases, and motor clusters spend an increasing amount of time passively cross-linking the MT-bundles while waiting for ATP binding [47]. In the limit of zero ATP, kinesin clusters passively crosslink the filaments, yielding a solid-like material [42]. These considerations hinder the simple interpretation of how the instability wavelength depends on the activity because tuning ATP concentration changes both the activity and elasticity.

To overcome the above-mentioned complexities, we titrated the concentrations of motor clusters at saturating ATP concentrations. The instability wavelength decreased sharply with increasing motor concentration (**Fig. 3B**). Second, we repeated the same experiment with clusters of non-processive K365 motors that detach from the MT after each step. We again observed decreasing wavelength with increasing motor concentration (**Fig. 3B**). This is consistent with an increase of the activity α. To explore the influence of elasticity, we kept the kinesin concentration constant while adding a varying amount of PRC-1, a specific crosslinker that bundles antiparallel microtubules while still allowing for their relative sliding [37,49]. Generally, increasing the number of crosslinkers should lead to stiffer bundles [50]. In agreement with this expectation, we find that increasing PRC-1 concentration increased the instability wavelength (**Fig. 3C**). As we did not change the amount of stress-generating motors, this is consistent with an increase of the sample elasticity $\kappa$. Taken together, these experiments confirm that in the highly confined regime, the instability wavelength is determined by the balance of active and elastic stresses, which are respectively controlled by the concentration of stress generating motors and non-motor crosslinkers.

To summarize, above a critical concentration of molecular motors, aligned 3D suspensions of extensile MT-bundles exhibit an active-stress driven hydrodynamic instability. The fastest-growing mode has a finite wavelength that is determined by confinement geometry and material properties. The observed scaling with the confinement size is consistent with a minimal hydrodynamic model, which predicts that the instability wavelength will be infinite in the bulk limit. However, reaching this limit is an experimental challenge. For square channels that are much larger than the critical length, the instability wavelength grows slower than the system size: $\lambda \propto \sqrt{W}$. Such weak scaling indicates that for all accessible regimes the active fluid structure is determined by the boundaries. We consider our findings in the context of the previously published results. First, we note that the instability wavelength of initially shear-aligned fluids is related to the velocity-velocity correlation length once the chaotic isotropic dynamics is developed. Notably, previous measurements of the correlation length were performed in microscopy chambers with essential unlimited lateral dimensions but strongly confined (~100 μm) axial direction [9,36]. Results shown here indicate that such measurements were influenced by the boundaries and therefore not representative of bulk dynamics. Second, in agreement with previous studies [9,36], the length scale of confined and dilute 3D isotropic fluids exhibited weak ATP dependence, in contrast to dense 2D MT active nematics [32,48]. Third, the ability of 3D active fluids to adjust their internal structure in response to changing boundaries provides new insights required for understanding the scale-invariant criterion that determines the onset of long-range transport flows in confined active fluids [23]. Taken together, our combined experimental and theoretical findings advance our understanding of both the pathways to low Reynold number 3D active turbulence and how such pathways are affected by boundaries [13].

Figures:

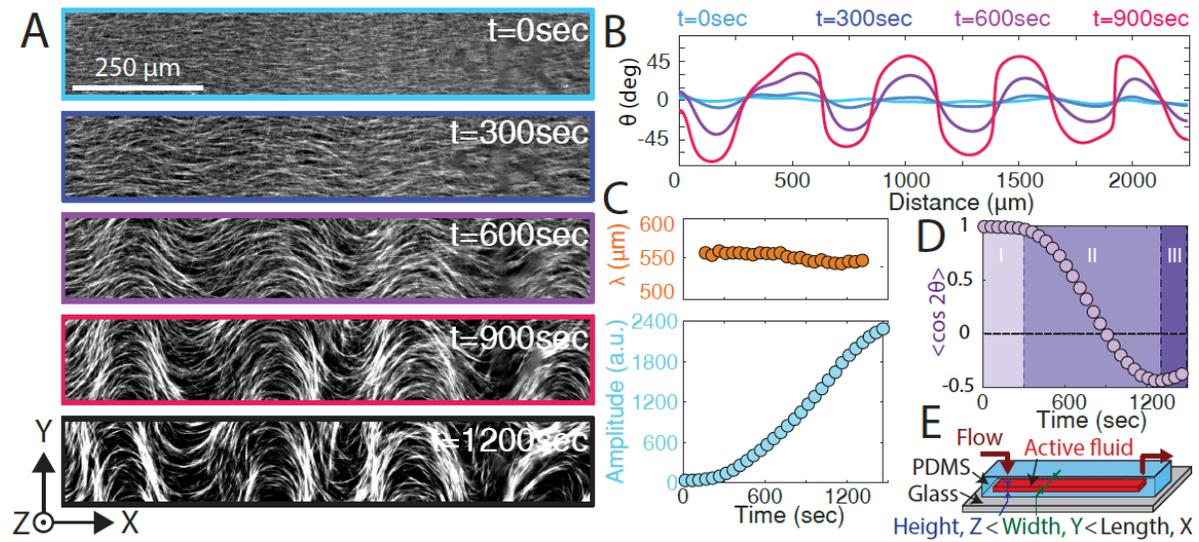

**Fig. 1: Bend instability in confined microtubule-based active nematics.** A) Instability of the flow-aligned 3D microtubule-based active isotropic fluid. B) Temporal evolution of the nematic director along the channel direction. θ is the angle between the local nematic director and the channel's long axis (x-axis). C) Instability wavelength and amplitude as a function of time. D) Temporal evolution of the alignment parameter $q=<\cos(2\theta)>$; $q$ quantifies the degree of ordering within a field of view. E) Schematic of the microfluidic channels with rectangular cross-sections. $H$=100 μm, $W$=3 mm, [KSA]=10 nM.

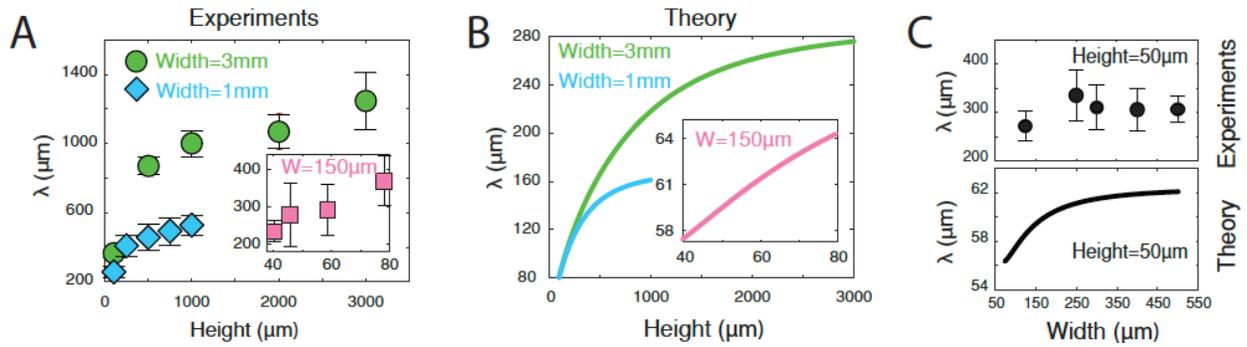

**Fig. 2: Channel's dimensions set the instability wavelength.** A) The instability wavelength increases with increasing channel's height. The width is always larger than or equal to the channel height. Inset: Wavelength increases linearly with thickness in the highly confined regime, $W=150$ μm. [KSA]=120 nM. B) Wavelength of the fastest growing mode as a function of channel height for the same channel width as in A obtained from the linear stability analysis for $\ell_\alpha = 40$ μm. Inset: fastest-growing mode for a width of 150 μm for the same value of $\ell_\alpha$. C) Wavelength as a function of channel height in the highly confined regime ($H=50$ μm, [KSA]=20 nM, $\ell_\alpha = 40$ μm).

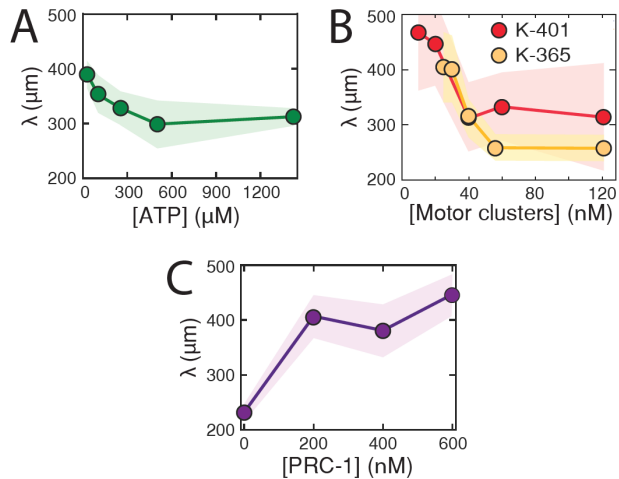

**Fig. 3: Molecular control of the instability wavelength in the highly confined regime.** Dependence of the instability wavelength for varying concentrations of A) Adenosine-5'-triphosphate (ATP), B) molecular motor clusters (processive K-401 and non-processive K-365) and C) MT-crosslinkers PRC-1. $W=3$ mm, $H=100$ μm.

**Acknowledgements:** We thank Bezia Lemma, Radhika Subramanian and Marc Ridilla for their help in protein purification. **Funding:** P. C. and Z.D. acknowledge support of Department of Energy – Office of Basic Energy Sciences under award DE-SC0019733. We also acknowledge the use of the optical, microfluidics, and biomaterial facilities supported by NSF MRSEC DMR-2011846. G.D, M.V., S.A.A. and A.B. acknowledge support of NSF MRSEC DMR-2011846. G.D acknowledges support from Brandeis University. A.B acknowledges funding from US-Israel Binational Science Foundation through BSF-2014279.

# Confinement Controlled Bend Instability of Three-Dimensional Active Fluids

Pooja Chandrakar[1,2], Minu Varghese[1], S.Ali Aghvami[1], Aparna Baskaran[1], Zvonimir Dogic[2,1], and Guillaume Duclos[1]

[1]Department of Physics, Brandeis University, Waltham, MA 02474
[2]Department of Physics, University of California at Santa Barbara, Santa Barbara, CA 93106

## Contents

## 1 Experimental Materials and Methods

**Microtubules (MTs)**: Tubulin dimers were purified from bovine brains through two cycles of polymerization - depolymerization in high molarity PIPES (1,4- piperazindiethanesulfonic) buffer [1]. Fluorophore-labeled tubulin was prepared by labeling the purified tubulin with Alexa-Fluor 647-NHS (Invitrogen, A-20006) [2]. GMPCPP (Guanosine 5'- ($\alpha$, $\beta$ methylenetriphosphate)), a non-hydrolyzable analogue of GTP was used to stabilize the dynamic instability in the MTs. Polymerization mixture consisted of 80 $\mu$M tubulin (with 3 % fluorescently labeled tubulin), 0.6 mM GMPCPP and 1mM DTT (dithiothreitol) in M2B buffer (80 mM PIPES, 1 mM EGTA, 2 mM MgCl2). After adding all the components, the mixture was incubated at 37 °C for 30 minutes, and subsequently for 6 hours at room temperature ($\sim$20 °C). This method resulted in MTs of $\sim$1.5 $\mu$m length [3]. MTs were aliquoted in small volumes (10 $\mu$L), flash-frozen in liquid nitrogen, and stored at -80 °C.

**Kinesin Motor Clusters and PRC1**: K401-BIO-6xHIS (processive motor, dimeric MW-110 kDa) and K365-BIO-6xHIS (non-processive motor, MW-50 kDa) are the 401 and 365 amino acids derived from N-terminal domain of Drosophila melanogaster kinesin-1, and labeled with 6-histidine and a biotin-tag [4][5]. The motor proteins were transformed and expressed in Rosetta (DE3) pLysS cells and purified following protocols described elsewhere [4]. The purified proteins were flash frozen in liquid nitrogen with 36% sucrose, and subsequently, stored in -80 °C. We used tetrameric streptavidin (ThermoFisher, 21122, MW: 52.8 kDa) to assemble clusters of biotin-labeled kinesins (KSA). To make K401-streptavidin clusters, 5.7 $\mu$L of 6.6 $\mu$M streptavidin was mixed with 5 $\mu$L of 6.4 $\mu$M K401 and 0.5 $\mu$L of 5 mM DTT in M2B. This mixture was incubated on ice for 30 minutes. K365-streptavidin clusters were prepared by mixing, 5.7 $\mu$L of 6.6 $\mu$M streptavidin, 3.1 $\mu$L of 20 $\mu$M K365, 0.5 $\mu$L of 5 mM DTT and 1.94 $\mu$L of M2B, and then left to incubate on ice for 30 minutes. PRC1 was used in the experiment to crosslink MTs while still allowing the relative sliding due to kinesin motors. PRC1 (MW: 72.5 kDa) was transformed and expressed in Rosetta BL21(DE3) cells, and subsequently purified as previously described [6].

**Preparation of Active Fluids**: The major components of the active fluid were MTs, kinesin-streptavidin clusters, ATP and a depletant Pluronic (F-127, Sigma P2443. MW: 12.5 kDa). An ATP regeneration system, consisting of phosphoenol pyruvate (PEP, Beantown Chemical, 129745) and pyruvate kinase/lactate dehydrogenase enzymes (PK/LDH, Sigma, P-0294), was used to maintain a constant concentration of ATP throughout the experiment. To minimize the fluorophore photobleaching during fluorescence microscopy, we used an oxygen scavenging system, consisting of glucose, DTT, glucose oxidase (Sigma, G2133) and catalase (Sigma, C40). The basic active fluid mixture consisted of 5.5 mM DTT, 3.3 mg/ml glucose, 0.22 mg/ml glucose oxidase, 0.039 mg/ml catalase, 2 mM trolox (Sigma, 238813), 26.67 mM PEP, 1.7 $\mu$L of PK/LDH, 3.29 mM of high-salt buffer - MgCl2 in M2B, 0.42% tracer particles (Polysciences, 18861), 1.42 mM ATP, 2% pluronic, 13.3 $\mu$M MTs, 121 nM (streptavidin conc) of K401 or K365 cluster. Samples were prepared fresh for each set of experiments. For experiments with varying PRC-1 concentration, the basic active fluid was doped with PRC-1 ranging from 200-600 nM.

**Microfluidic Channels**: For channels with different height (Fig. 2), a molding master was created by machining cyclic olefin copolymer (COC) [7]. Channels were then replicated in Poly (dimethyl siloxane) (PDMS). Once PDMS was cured at 70 °C for 1 hour, it was removed from the master and holes for inlet and outlet were punched. Afterwards, the PDMS channel was plasma-bonded to a glass slide. The channels were passivated with 2% Pluronic solution for 30 minutes, to avoid protein adsorption onto the glass and PDMS surfaces.

Microfluidic channels of varying width were microfabricated using standard soft photolithography techniques. We built-in pneumatic quake valves [8] to prevent any external flows once the microtubules were shear-aligned. We used "push-up" type of valves, in which flow channels are located above the control channels.

**Time-Lapse Microscopy**: Time-lapse experiments were performed in fluorescence on a widefield microscope (Nikon TE-2000) equipped with a Prior XY stage, and a 4X, 0.15 NA objective. The displacement of the sample, the illumination (Lumencor Sola) and the acquisition (Clara E, Andor) were controlled by micro-manager, a software package for controlling automated microscopes [9]. The typical delay between two successive images of the same FOV was set from 500 ms to 10 sec, depending on the ATP and KSA concentrations. When possible, multi positions were acquired within a same sample. For the results shown in Fig. 2A, 3 $\mu$m diameter passive fluorescent particles (Polysciences, 18861) were homogeneously dispersed in the active fluid and imaged on a Nikon Ti-E microscope with a 1X Objective. The acquisition was controlled by micro-manager.

**Confocal Microscopy**: We used confocal microscopy to image slices of fluids with the weakest confinement (2 mm height) to check the initial microtubule bundle alignment (supplementary movie S5). The experiment was performed on an inverted microscope (Nikon, Ti2- Eclipse) equipped with spinning disk module (CrestoOptics, X-Light V2) and an automated XY stage (ASI MS2000). Fluorescently-labeled MTs were illuminated with the appropriate excitation wavelength from a light source (lumencor, CELESTA light engine). A Hamamatsu camera (Orca 4.0 V1) and a 40X objective (Nikon Apo LWD) was used for imaging the samples. The step size was set to 2.6 $\mu$m. For SI Fig. 5, a Leica SP8 with a 20X objective and a step size of 5 $\mu$m was used.

**Image Processing**:

**Orientation**: Most of the image processing was performed using the ImageJ public domain software [10] and Matlab (MathWorks) [11]. The orientation field was obtained by computing the local 2D structure tensor with the ImageJ plugin OrientationJ [10], [12]. The window size was set to 20 pixels = 32.3 $\mu$m.

**Velocity**: The velocity field was mapped by particle image velocimetry (PIV) analysis. Stacks of images were analyzed with a custom-made PIV algorithm based on the MatPIV software package for Matlab (MathWorks) [13]. The window size was set to 32 pixels = 51.6 $\mu$m with a 0.75 overlap. The velocity was measured either on timeseries of the microtubules or of the embedded particles. We checked that both channels gave similar velocities fields (supplementary movie S4)

**Measurement of the instability wavelength and amplitude:** The amplitude and wavelength of the instability were measured from heatmaps of the velocity fields along the Y-axis (direction of instability growth). We integrated the displacement fields over time to measure the instability amplitude. For each Field Of View (FOV) and each time frame, Fast Fourier Transform of the displacement profiles was performed with Matlab along the channel direction for each row of the FOV. We made sure to capture at least 3 wavelengths to restrict any finite-size bias. We measured both the amplitude of the FFT largest mode and its wavelength. Both quantities were then averaged over the FOV.

The following number of instability cycles (N) were analyzed per experimental condition, from at least two replicas:

Effect of channel size on the instability wavelength (Fig. 2):
(W = 1 mm: H = 100 $\mu$m, N = 8; H = 250 $\mu$m, N = 9; H = 500 $\mu$m, N = 10; H = 750 $\mu$m, N = 14; H = 1 mm, N = 7).
(W = 150 $\mu$m: H = 40.5 $\mu$m, N = 47; H = 45.8 $\mu$m, N = 61; H = 58.6 $\mu$m, N = 47; H = 78 $\mu$m, N = 33).
(W = 3 mm: H = 100 $\mu$m, N = 14; H = 500 $\mu$m, N = 14; H = 1 mm, N = 16; H = 2 mm, N = 19; H = 3 mm, N = 7).
(H = 100 $\mu$m: W = 125 $\mu$m, N = 31; W = 250 $\mu$m, N = 25; W = 300 $\mu$m, N = 15; W = 400 $\mu$m, N = 13; W = 500 $\mu$m, N = 15).

Effect of motors, crosslinkers and ATP concentrations on the instability wavelength (Fig. 3):
([ATP] = 25 $\mu$M, N = 5; [ATP] = 100 $\mu$M, N = 5; [ATP] = 250 $\mu$M, N = 5; [ATP] = 500 $\mu$M, N = 5; [ATP] = 1420 $\mu$M, N = 5).
([K401] = 10 nM, N = 25; [K401] = 20 nM, N = 20; [K401] = 40 nM, N = 29; [K401] = 60 nM, N = 32; [K401] = 120 nM, N=86).
([K365] = 25 nM, N = 36; [K365] = 30 nM, N = 24; [K365] = 40 nM, N = 32; [K365] = 56 nM, N = 32; [K365] = 120 nM, N = 18).
([PRC1] = 0 nM, N = 6; [PRC1] = 200 nM, N = 14; [PRC1] = 400 nM, N = 4; [PRC1] = 600 nM, N = 3);

## 2 Theoretical Model

**Hydrodynamic Model**: Since the microtubule bundles have head-tail symmetry and are invariant under rotations about their long axes, we can describe their orientational order using a uniaxial nematic order parameter, $\mathbf{Q} = S\left(\hat{n}\hat{n} - \frac{I}{3}\right)$, where $S$ is the magnitude of order and $\pm\hat{n}$ is the mean axis of orientation. In the presence of an ambient flow $\vec{u}$, the orientation of microtubule bundles evolve with time as [14, 15]:

$$\partial_t \mathbf{Q} + \vec{u}.\vec{\nabla}\mathbf{Q} + \mathbf{Q}.\mathbf{\Omega} - \mathbf{\Omega}.\mathbf{Q} = D_R(-a_2 + \kappa\nabla^2)\mathbf{Q} + \frac{2}{3}\xi\mathbf{E} \qquad (1)$$

where $\Omega_{ij} = \frac{1}{2}(\partial_j u_i - \partial_i u_j)$, and $E_{ij} = \frac{1}{2}(\partial_j u_i + \partial_i u_j)$. The left hand side of equation (1) is the comoving, corotational time derivative of $\mathbf{Q}$. The first term on the right hand side captures that in the absence of aligning flows, the active fluid settles into a uniform isotropic state ($\mathbf{Q} = 0$) at a rate determined by the rotational diffusion constant, $D_R$. $a_2$ is a phenomenological constant that depends on the concentration of microtubule bundles, and $\kappa$ is the constant of elasticity. The last term accounts that the shear flows align the microtubule bundles, due to their elongated shape. The flow alignement parameter $\xi$ depends on the aspect ratio of the microtubule bundles, $h$, through $\xi = \frac{h^2-1}{h^2+1}$ [16]. Since the microtubule bundles are long and thin, $h \to \infty$ and $\xi \to 1$. We have ignored higher order flow alignment terms for simplicity. Including them doesn't qualitatively change the results presented below.

When microtubule bundles elongate outward, they exert a dipolar stress on the fluid, $-\alpha \mathbf{Q}$ [17], where $\alpha > 0$ is the magnitude of the force dipole associated with a single bundle. Since the Reynolds number associated with the active fluid is small ($\sim 10^{-6}$), the flows arise from an instantaneous balance of the viscous and active forces:

$$\vec{\nabla}.(\eta \mathbf{E} - p\mathbf{I} - \alpha \mathbf{Q}) = 0$$
$$\Rightarrow \eta \nabla^2 \vec{u} - \vec{\nabla} p - \alpha \vec{\nabla}.\mathbf{Q} = 0 \tag{2}$$

where $\eta$ is the viscosity and $p$ is a mechanical pressure arising from incompressibility of the fluid,

$$\vec{\nabla}.\vec{u} = 0 \tag{3}$$

In the following discussion, we analyze the stability of equations 1-3, starting from a uniform orientationally ordered stationary state. We show that there are two unstable eigenmodes: one consists of twist and bend, and the other is a combination of splay, bend, and orientational order.

**Linear Stability Analysis**: After flow aligning the microtubules along $\hat{x}$, the orientational order is given by $\mathbf{Q}_0 = S_0 \left( \hat{x}\hat{x} - \frac{1}{3} \right)$ and $\vec{u} = 0$. $\mathbf{Q}_0$ is not a steady state, because the orientational order decays over time as $S_0 e^{-D_R a_2 t}$. However, the timescale for this decay, of order $1/D_R a_2$, is large compared to the active timescale, $\eta/\alpha$. This is the reason why the experimental system doesn't relax to the isotropic state as soon as the shear flow is stopped. Instead it exhibits the activity-driven instability. Since we study dynamics on time scales $t \ll 1/D_R a_2$, in the following discussion we neglect the exponential decay in the magnitude of the nematic order.

Consider a small perturbation in orientational order, such that $\mathbf{Q} = \mathbf{Q}_0 + \delta\mathbf{Q}$. The evolution of this perturbed state, upto linear order is given by:

$$\partial_t \delta \mathbf{Q} + \mathbf{Q}_0.\delta\mathbf{\Omega} - \delta\mathbf{\Omega}.\mathbf{Q}_0 = D_R \kappa \nabla^2 \delta\mathbf{Q} + \frac{2}{3}\xi \delta\mathbf{E}. \tag{4}$$

Let $\delta\mathbf{Q} = S_0 \delta S \left( \hat{x}\hat{x} - \frac{I}{3} \right) + S_0 (\delta\vec{n}_\perp \hat{x} + \hat{x}\delta\vec{n}_\perp)$, where $\delta \hat{n}_\perp = (\delta n_y, \delta n_z)$. Consider the perturbations to be of the form $\delta a = \delta \tilde{a}\, e^{-i\vec{k}.\vec{r}}$. Then,

$$\partial_t \delta \tilde{S} = -D_R \kappa k^2 \delta \tilde{S} - \frac{i\xi}{S_0} k_x \delta \tilde{u}_x \tag{5}$$

$$\partial_t \delta \tilde{\vec{n}}_\perp = -D_R \kappa k^2 \delta \tilde{\vec{n}}_\perp + \frac{iS_0^-}{S_0} \vec{k}_\perp \delta \tilde{u}_x - \frac{iS_0^+}{S_0} k_x \delta \tilde{\vec{u}}_\perp \tag{6}$$

where $S_0^- = (S_0 - \xi)/2$, and $S_0^+ = (S_0 + \xi)/2$. From equations (2) and (3),

$$\delta \tilde{u}_x = \frac{i\alpha S_0}{\eta k^4} \left[ (k_\perp^2 - k_x^2)(\vec{k}_\perp.\delta\tilde{\vec{n}}_\perp) + k_x k_\perp^2 \delta \tilde{S} \right] \tag{7}$$

$$\delta \tilde{\vec{u}}_\perp = \frac{i\alpha S_0}{\eta k^4} \left[ k^2 k_x \delta \tilde{\vec{n}}_\perp - 2k_x \vec{k}_\perp (\vec{k}_\perp.\delta\tilde{\vec{n}}_\perp) - k_x^2 \vec{k}_\perp \delta \tilde{S} \right] \tag{8}$$

Eliminating the flow fields yields:

$$\partial_t \delta \tilde{S} = \left(-D_R \kappa k^2 + \frac{\xi \alpha}{\eta k^4} k_x^2 k_\perp^2\right) \delta \tilde{S} + \frac{\xi \alpha}{\eta k^4} k_x (k_\perp^2 - k_x^2)(\vec{k}_\perp . \delta \tilde{\vec{n}}_\perp) \tag{9}$$

$$\partial_t \delta \tilde{\vec{n}}_\perp = -D_R \kappa k^2 \delta \tilde{\vec{n}}_\perp - \frac{\alpha S_0^-}{\eta k^4} \vec{k}_\perp \left[(k_\perp^2 - k_x^2)(\vec{k}_\perp . \delta \tilde{\vec{n}}_\perp) + k_x k_\perp^2 \delta \tilde{S}\right]$$
$$+ \frac{\alpha S_0^+}{\eta k^4} k_x \left[k^2 k_x \delta \tilde{\vec{n}}_\perp - 2 k_x \vec{k}_\perp (\vec{k}_\perp . \delta \tilde{\vec{n}}_\perp) - k_x^2 \vec{k}_\perp \delta \tilde{S}\right] \tag{10}$$

**Twist-bend and splay-bend-order modes**: When $k_\perp \neq 0$, the dynamics of $\delta \tilde{S}$, $\delta \tilde{n}_y$, and $\delta \tilde{n}_z$ are coupled. In this case, one of the eigenmodes is twist, and the other two eigenmodes are combinations of splay and orientational order. To fist order in the perturbations, splay deformations are given by $\vec{k}_\perp . \delta \tilde{\vec{n}}_\perp$, and twist deformations by $\hat{x}.(\vec{k} \times \delta \tilde{\vec{n}}_\perp)$.
From eqns. (9) and (10),

$$\partial_t [\hat{x}.(\vec{k} \times \delta \tilde{\vec{n}}_\perp)] = \left(-D_R \kappa (k_x^2 + k_\perp^2) + \frac{\alpha S_0^+}{\eta} \frac{k_x^2}{k_x^2 + k_\perp^2}\right) [\hat{x}.(\vec{k} \times \delta \tilde{\vec{n}}_\perp)] \tag{11}$$

and

$$\partial_t \begin{pmatrix} \delta \tilde{S} \\ \vec{k}_\perp . \delta \tilde{\vec{n}}_\perp \end{pmatrix} = \begin{pmatrix} -D_R \kappa k^2 + \frac{\xi \alpha}{\eta k^4} k_x^2 k_\perp^2 & \frac{\xi \alpha}{\eta k^4} k_x (k_\perp^2 - k_x^2) \\ -\frac{\alpha}{\eta k^4} k_x k_\perp^2 (S_0^+ k_x^2 + S_0^- k_\perp^2) & -D_R \kappa k^2 + \frac{\alpha}{\eta k^4} (S_0^+ k_x^2 + S_0^- k_\perp^2)(k_x^2 - k_\perp^2) \end{pmatrix} \begin{pmatrix} \delta \tilde{S} \\ \vec{k}_\perp . \delta \tilde{\vec{n}}_\perp \end{pmatrix}$$

The above set of equations has one unstable eigenmode:

$$\partial_t [-\xi k_x \delta \tilde{S} + (S_0^+ k_x^2 + S_0^- k_\perp^2)(\vec{k}_\perp . \delta \tilde{\vec{n}}_\perp)] = \left[-D_R \kappa (k_x^2 + k_\perp^2) + \frac{\alpha}{\eta} \frac{(k_x^4 S_0^+ - k_\perp^4 S_0^-)}{(k_x^2 + k_\perp^2)^2}\right] \times$$
$$[-\xi k_x \delta \tilde{S} + (S_0^+ k_x^2 + S_0^- k_\perp^2)(\vec{k}_\perp . \delta \tilde{\vec{n}}_\perp)] \tag{12}$$

Since bend deformations are given by $\hat{x} \times (\vec{k} \times \delta \tilde{\vec{n}}_\perp) = -k_x \delta \tilde{\vec{n}}_\perp$, the growth of $\delta \tilde{n}_y$ or $\delta \tilde{n}_z$ at nonzero $k_x$ must involve a bend. Therefore, eqn. (11) is in fact a twist-bend mode, and eqn. (12) is a splay-bend-orientational order mode.

**Bulk limit**: The growth rate of both modes approach their maximum possible value when $k_\perp \to 0$ (see SI Fig. 8a). In this case, both the unstable modes correspond to pure bend. The dynamics of $\delta \tilde{n}_y$ and $\delta \tilde{n}_z$ decouple, and are given by:

$$\partial_t \delta \tilde{\vec{n}}_\perp = \left(-D_R \kappa k_x^2 + \frac{\alpha S_0^+}{\eta}\right) \delta \tilde{\vec{n}}_\perp \tag{13}$$

The nematic order remains uniform since $\delta \tilde{S} = -D_R \kappa k^2 \delta \tilde{S}$. Thus, the fastest growing deformation in a bulk orientationally ordered extensile active fluid is a long wavelength pure bend.

**Dominance of twist-bend mode**: At nonzero $k_\perp$, the activity needs to be larger than a critical value for the existance of any instability. Also, the instabilities are no longer scale-free. In each case, increasing $k_\perp$ increases the value of $k_x$ at which the growth rate is maximum, while decreasing the growth rate associated with this fastest growing wavemode(see SI Fig. 8a).

In the ordered regime ($S_0 > \xi$), the growth rate of the twist-bend mode is always larger than the growth rate of the splay-bend-order mode. For $S_0 < \xi$, splay-bend-order is more unstable than twist-bend in the limit $k_x \to 0$. If $S_0$ is sufficiently small, the growth rate of the $k_x = 0$ splay-order mode is larger than that of the fastest growing twist-bend mode. In this case, a spontaneous flow instability precedes the bend instability (see SI Fig. 8b). Since the experiments are performed on a highly ordered state, the emergent properties seen in the experiment will be those associated with the twist-bend mode.

**Effect of confinement**: From eqn. (11), the growth rate of the twist-bend mode is

$$\nu_{tb} = -D_R \kappa (k_x^2 + k_\perp^2) + \frac{\alpha(S_0 + \xi)}{2\eta} \frac{k_x^2}{k_x^2 + k_\perp^2} \quad (14)$$

The presence of confining boundaries in our system implies that $k_\perp$ has to be non-zero. For instance, due to no-slip boundary conditions at the walls, the smoothest possible perturbations about the stationary state $\sim \sin(\pi y/W)\sin(\pi z/H)$, where the width $W$ is the distance between the walls in the y direction, and the height $H$ is the distance between the walls in the z direction. Therefore, the minimum value of $k_\perp$ is $\sqrt{\frac{\pi^2}{W^2} + \frac{\pi^2}{H^2}}$.

When the confinement lengthscale, $\pi/k_\perp$, is larger than the active lengthscale, $\ell_\alpha = 2\pi\sqrt{\frac{2\eta D_R \kappa}{\alpha(S_0+\xi)}}$, there is a finite range of values of $k_x$ for which $\nu_{tb} > 0$. The fastest growing mode corresponds to

$$k_x^2 = \frac{2\pi}{\ell_\alpha} k_\perp - k_\perp^2$$

This mode grows at a rate

$$\nu^{\max} = -D_R \kappa \frac{2\pi}{\ell_\alpha} k_\perp + \frac{\alpha(S_0+\xi)}{2\eta}\left(1 - \frac{\ell_\alpha}{2\pi} k_\perp\right)$$

Using $k_x = 2\pi/\lambda$, and $k_\perp = \sqrt{\frac{\pi^2}{W^2} + \frac{\pi^2}{H^2}}$,

$$\lambda = 2\left[\frac{2}{\ell_\alpha}\left(\frac{1}{W^2} + \frac{1}{H^2}\right)^{\frac{1}{2}} - \left(\frac{1}{W^2} + \frac{1}{H^2}\right)\right]^{-\frac{1}{2}} \quad (15)$$

and

$$\nu^{\max} = \frac{\alpha(S_0+\xi)}{2\eta} - 2\pi\sqrt{\frac{\alpha(S_0+\xi)D_R\kappa}{2\eta}}\sqrt{\frac{1}{W^2} + \frac{1}{H^2}} \quad (16)$$

The twist-bend mode itself is given by $\hat{x}\cdot(\vec{k}\times\delta\tilde{\vec{n}}_\perp) = \delta\tilde{n}_z/W - \delta\tilde{n}_y/H$. If $H \ll W$, the unstable eigenmode consists predominantly of $\delta\tilde{n}_y$. Therefore, the instability will appear to be in the xy plane.

In the minimal hydrodynamic model presented here, we have ignored spatial variations in the concentration of the microtubule bundles. If these variations were included, concentration fluctuations couple to splay and order fluctuations, changing the second unstable eigenmode into a splay-bend-order-concentration mode. The lowest order effect of activity on concentration is to induce a flux from regions of high speed to regions of low speed. This could be the reason why concentration and velocity are anticorrelated in the experiments. Our calculation also ignored passive stresses. Including passive stresses in the analysis increases the value of $\ell_\alpha$, but doesn't change the functional dependence of the fastest growing wavemode on the channel dimensions.

# 3 Supplementary Figures

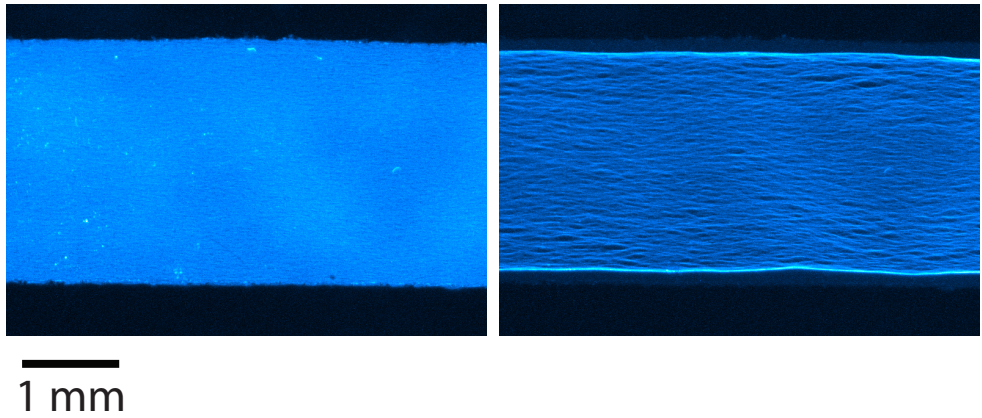

SI Fig. 1: **Experimental criterion for no instability**. A critical concentration of kinesin clusters is required for the instability to grow. No instability has been observed when motor cluster concentration is less than 5 nM. Instead, the whole sample contracts, as previously reported in [18] ([ATP] = 1.4 mM, [KSA] = 5 nM, W = 2.5 mm, H = 100 $\mu$m, ). See supplementary Movie S3.

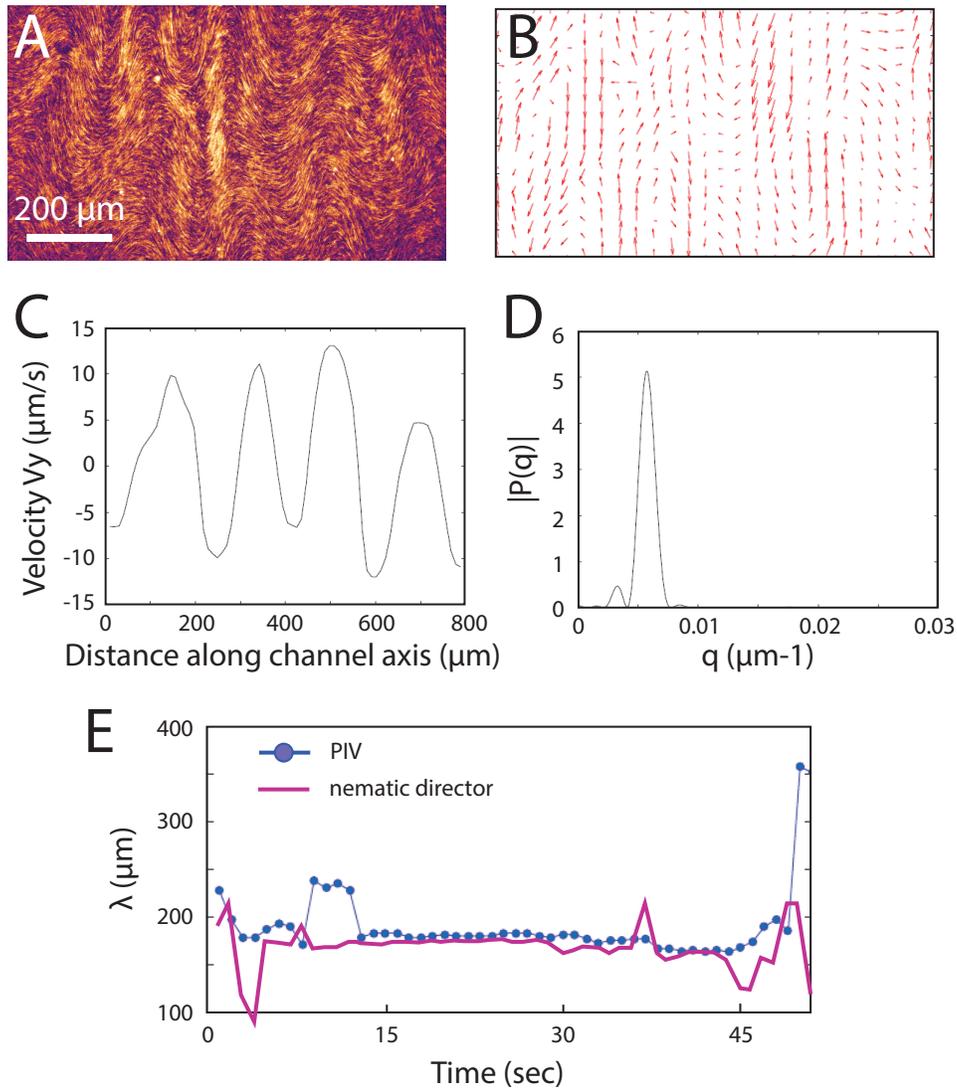

SI Fig. 2: **Measurement of the instability wavelength** . A) Fluorescent image of the microtubule bundles during the growth of the instability. [ATP] = 1.42 mM, [KSA] = 120 nM. B) Corresponding velocity field measured by Particle Image Velocimetry. C) Y-component of the velocity, for a fixed y, along the channel axis (x-direction). D) FFT spectrum of the velocity profile (panel C). E) Temporal evolution of the instability wavelength (averaged over the field of view), measured from the velocity field (blue) and the orientation field (pink). [ATP] = 1.4 mM, [KSA] = 120 nM, W = 3 mm, H = 100 $\mu$m.

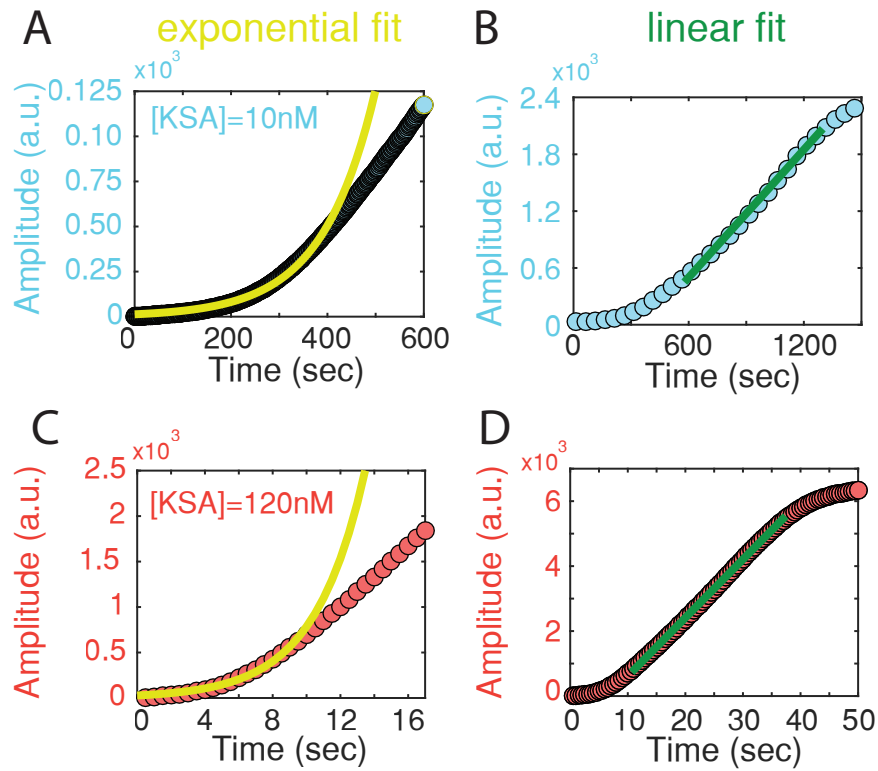

SI Fig. 3: **Temporal evolution of the instability amplitude for two different concentrations of motor clusters**. The instability first grows exponentially and later transitions to a linear regime. A-B) Amplitude for 10 nM of motor clusters (KSA), C-D) 120 nM of motor clusters (KSA). [ATP] = 1.4 mM, W = 3 mm, H = 100 $\mu$m.

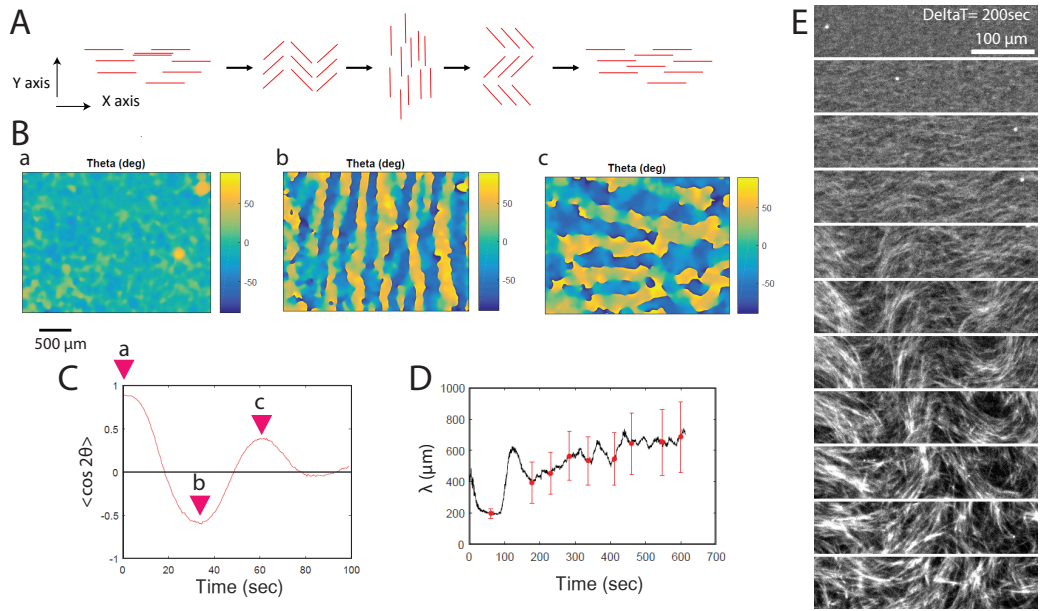

SI Fig. 4: **Cascade of orthogonal instabilities.** Noise in the direction of propagation of instabilities accumulate over time and the active fluid evolves in a chaotic state with no memory of the initial long-range order. A) Schematic of the successive instabilities. B) heatmap of the orientation field $\theta$ for q extrema: (a) q $\sim$ 1 (Initial state), (b) q $\sim$ -0.5 and (c) q $\sim$ 0.25. C) Temporal evolution of the alignment parameter q = $\langle \cos(2\theta) \rangle$ over multiple orthogonal instabilities. D) Temporal evolution of the wavelength for multiple instabilities. The wavelength is increasing over time. [ATP] = 1.4 mM, [KSA] = 120 nM, W = 3 mm, H = 100 $\mu$m. E) Coarsening of the microtubule bundles over time. [ATP] = 1.4 mM, [KSA] = 10 nM, W = 3 mm, H = 100 $\mu$m.

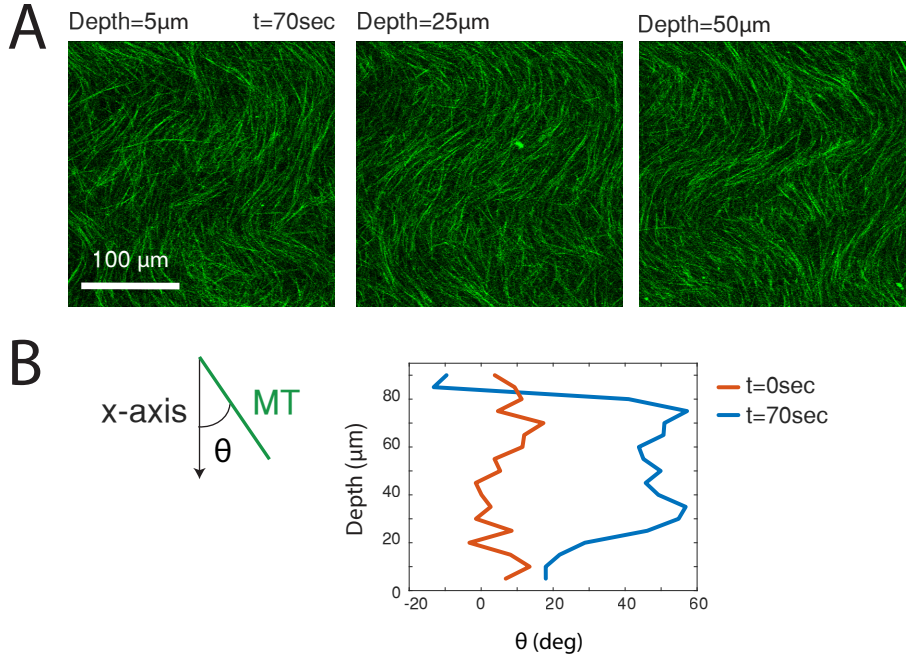

SI Fig. 5: **Orientation angle measurement of microtubule bundles at various depths** A) Confocal slices of the MT bundles at varying depths along the z-axis of the channel. The in-plane orientation of the MT bundless is not uniform at different z-planes. The bend is more pronounced in the mid-plane. B) Orientation of the MT bundles as a function of z-positions at t =0 sec and t = 70 sec, when the instability is well developed. [ATP] = 25 $\mu$M, [KSA] = 120 nM. W = 3 mm, H = 100 $\mu$m.

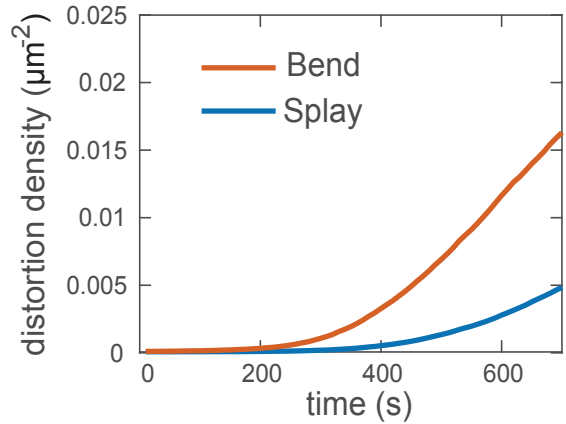

SI Fig. 6: **Temporal evolution of the splay (blue) and bend (red) distortion densities during the growth of the instability**. Bend - $\langle (\nabla \cdot \vec{n})^2 \rangle$ and splay - $\langle (\vec{n} \times (\nabla \times \vec{n}))_y^2 \rangle$ distortion densities measured from the in-plane orientation field ($\vec{n}$) of the MT-bundles. Bend deformations grow faster than splay deformations. [KSA] = 10 nM. W = 3 mm, H = 100 $\mu$m.

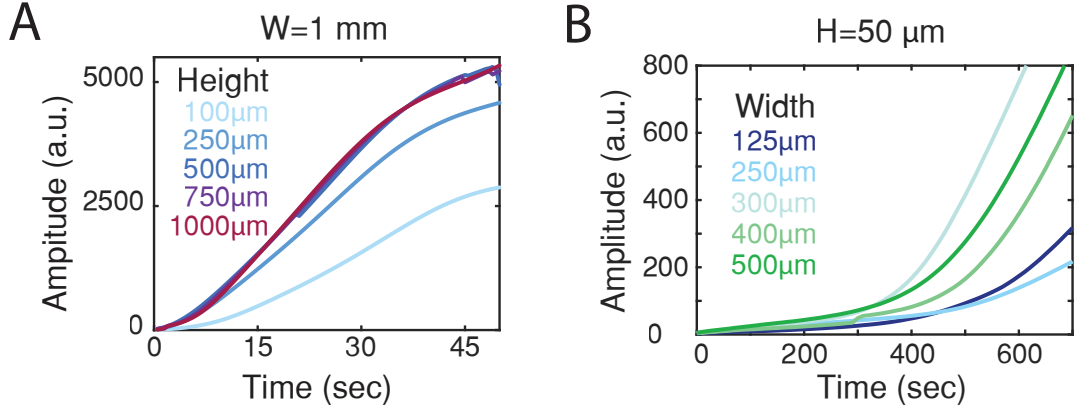

SI Fig. 7: **Temporal evolution of instability amplitude as a function of channels' width and height**. A) The growth rate of the instability increases when increasing the channel's height. (W = 1 mm, [KSA] = 120 nM, [ATP] = 1.4 mM). B) The growth rate of the instability does not show a clear dependence with the channel's width (H = 50 $\mu$m, [KSA] = 20 nM, [ATP] = 1.4 mM).

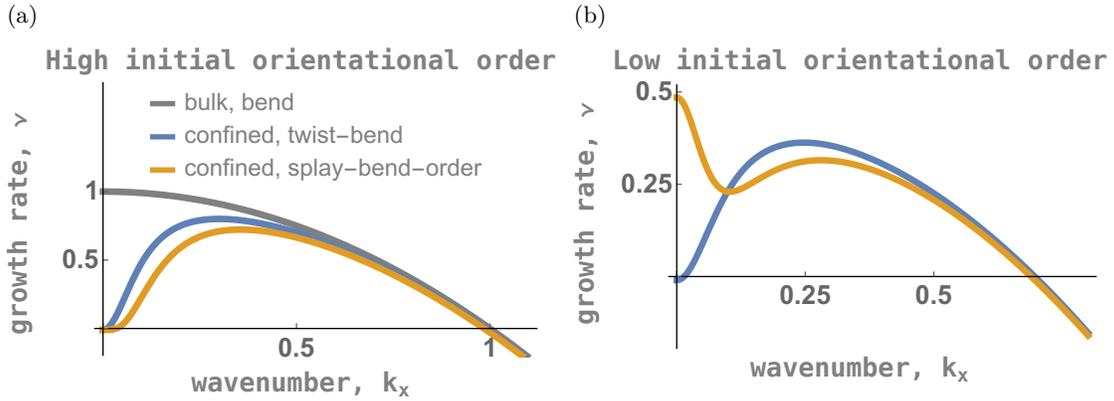

SI Fig. 8: **Theoretical prediction of the instability growth rate for bulk and confined systems.** (a) In a bulk orientationally ordered extensile active fluid, long wavelength pure bend modes have the fastest growth rate. When confined, there are two distinct unstable modes: twist-bend (blue) and splay-bend-orientational order (orange). When the initial orientational order is high ($S_0 > \xi$), growth rate of the twist-bend mode is always larger than that of the splay-bend-order mode. The fastest growing mode is a twist-bend with a finite wave number. (b) When the initial orientational order is sufficiently low ($S_0 \ll \xi$), the fastest growing mode is a uniform ($k_x = 0$) splay-bend-order mode. In this case, a spontaneous flow instability supersedes the twist-bend instability, and can give rise to long range coherent flows [19]. For all plots, $D_R \kappa = 1$, $\frac{\alpha}{\eta} = 1$, and $\lambda = 1$. Confinement was incorporated by setting $k_\perp = 0.01$, while the bulk limit corresponds to $k_\perp = 0$. For high initial order $S_0 = 1$, while for low initial order $S_0 = 0.1$.

## 4 Supplementary Movie Captions

**Movie S1** Instability in an orientationally ordered, MT-based active fluid. The instability grows in the (x-y) plan, perpendicularly to the thinnest dimension axis (z-axis). [KSA] = 10 nM, [ATP] = 1.4 mM, H = 100 $\mu$m and W = 3 mm.

**Movie S2** Instability in an orientationally ordered, MT-based active fluid at a higher motor cluster concentration than the active fluid in Movie S1. [KSA] = 60 nM, [ATP] = 1.4 mM, H = 100 $\mu$m and W = 3 mm. The instability grows faster and has a shorter wavelength at higher motor concentration.

**Movie S3** Instability is suppressed below a critical motor concentration (5 nM of motor cluster). [KSA] = 5 nM, [ATP] = 1.4 mM, H = 100 $\mu$m and W = 3 mm.

**Movie S4** Instability observation via fluorescently labeled MTs (upper panels), and tracer particles (lower panels) and respective Y-component of the velocity field.

**Movie S5** Z-stack of an active fluid with no ATP after beein shear-aligned along the channel long axis (x-axis). The MT-bundles are strongly aligned along the x-axis at various depths. H = 2 mm and W = 3 mm. [KSA] = 120 nM, [ATP] = 0 mM.

**Movie S6** Instability in active fluids confined in channels of increasing heights: 100 $\mu$m, 500 $\mu$m, 1 mm and 2 mm (top to bottom panel). Width of the channel was kept constant - 3 mm. The color map shows the y-component of velocity measured by PIV. The wavelength of the instability strongly depends on the smaller channel dimension: the instability wavelength increases when increasing the channel height. [KSA] = 120 nM, [ATP] = 1.4 mM.

**Movie S7** Instability in active fluids confined in channels of various widths: 125 $\mu$m, 300 $\mu$m, 400 $\mu$m and 500 $\mu$m (top to bottom panel). Height of the channel was kept constant - 50 $\mu$m. In that regime, increasing the channel height does not influence the instability wavelength. [KSA] = 20 nM, [ATP] = 1.4 mM.